# Use of biogenic nanomaterials to improve the peritoneal dialysis technique: A Translational Research Perspective

**Dinesh Kumar**

*Centre of Biomedical Research (CBMR), SGPGIMS Campus, R B Road, Lucknow-226014, Uttar Pradesh, India*

**Address for Correspondence:**

**Dr. Dinesh Kumar**
(Assistant Professor)
Centre of Biomedical Research (CBMR),
Lucknow-226014, Uttar Pradesh, India
Mobile: +91-8953261506; +91-9044951791

**Email:** dineshcbmr@gmail.com
Webpage: https://dineshnmr.wordpress.com/

**Abstract:**

Intraperitoneal and catheter exit site infections are the most common complications associated with prolonged peritoneal dialysis (PD) therapy used for treating the patients with end stage renal failure (ESRF). Recurrent and persistent infections often cause inflammation of the peritoneum, a condition known as infectious peritonitis and to resolve the condition, patients require antibiotic treatment. However, if the treatment is delayed or if it fails due to antibiotic resistance, the peritonitis may lead to permanent malfunctioning of peritoneal membrane causing technique failure and transferring the patients to haemodialysis. Severe and prolonged peritonitis is not only the major cause of technique failure, it is also the leading cause of mortality and morbidity in PD patients. Therefore, there is an urgent need to improve the existing PD technique so that the frequency of PD associated infections could be reduced and infectious peritonitis episodes thereof during prolonged peritoneal dialysis. In this perspective, I highlight the possibility to improve the PD technique through the use of antimicrobial nanoparticles synthesized biologically.

## Perspective

Peritoneal dialysis (PD) is a well-established renal replacement therapy (RRT) used for treating patients with end-stage renal failure (ESRF) [1,2] **(Fig. 1A)**. It is highly cost-effective and offers certain clear advantages over haemodialysis (HD) such as simplicity, flexible lifestyle, reduced need for trained technicians and nurses, minimal technical support requirement, lack of electricity dependence, and home-based therapy [1,3-5]. Compared to HD, it also provides stable hemodynamics and better preservation of residual renal function [1,3-5]. Though PD has potential cost savings and offers several advantages over HD [6], however, it is often associated with a high risk of infection of the intraperitoneal cavity, subcutaneous tunnel and catheter exit site and subsequently formed microbial biofilms. After cardiovascular disease, infection is the second leading reason for admission to hospital among patients receiving long-term dialysis [7]. Mostly, the PD patients suffer from bacterial and fungal infections. Bacterial infections generally develop from colonization of sporadic microbial contaminations either through the stomach or through the catheter exit site and fungal infections may develop subsequent to antibiotic use. In most of the cases, the infections resolve through the use of empiric anti-biotic treatment. The prescribed anti-biotic treatment is well justified if the infections appear occasionally. However, ESRF patients continuing on PD generally suffer from inherently weak and sabotaged immune system; therefore the infections are as frequent as once after every 10-15 weeks. However, the major drawback associated with frequent use of conventional antimicrobial agents is the development of multiple drug resistance. Drug resistance enforces high antibiotic dose administration which may cause intolerable toxicity or adverse side effects.

Further, if the infections sustain for a week or more (either because of delayed treatment or multidrug resistant pathogens), these may lead to the inflammation of the peritoneum or abdominal cavity lining, the condition commonly known as infectious peritonitis [8,9]. The condition severely affects the functioning of peritoneal membrane and to resolve the condition, patients require hospitalization which not only poses financial burden on such patients because of higher medical costs, but, also poses a burden on nephrology wards for admitting such patients. Established treatment of infectious peritonitis involves rapid resolution of inflammation by eradicating the causative organism(s) and preservation of peritoneal membrane function. In severe cases, the intravenous antibiotics may become necessary. However, sometimes, even after the use of intravenous and intraperitoneal antibiotics, the treatment may fail to resolve the infectious peritonitis and the patients are transferred to haemodialysis (HD), either temporarily or permanently; thus reducing patient's autonomy and increasing hospitalization [2,9].

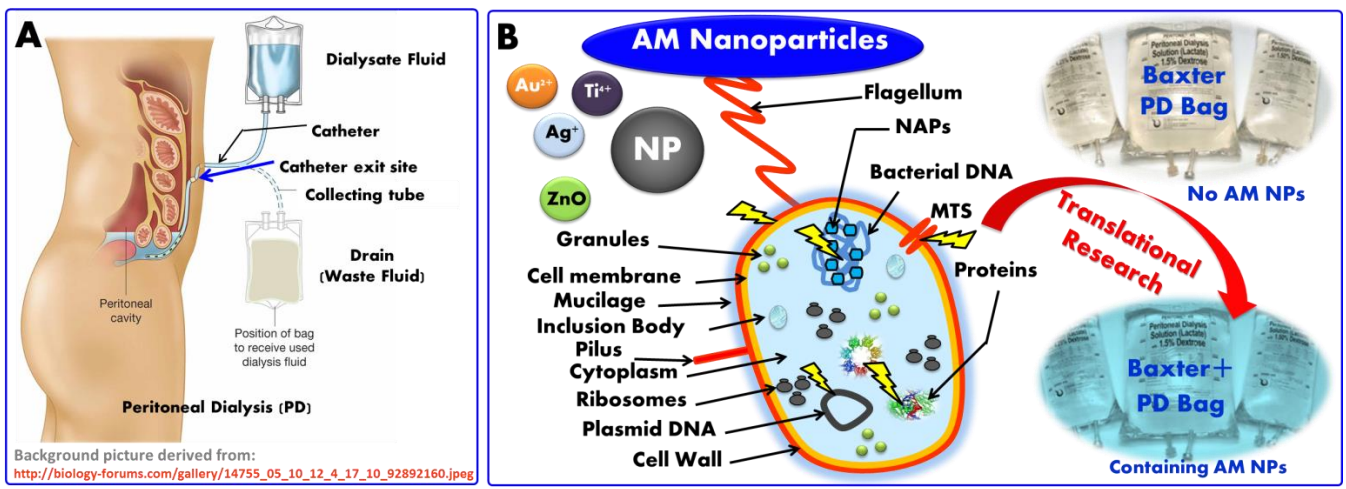

*Figure 1: (A)* Pictorial representation of peritoneal dialysis procedure and *(B)* schematic showing the generalized mechanism of biocides activity of AM nanoparticles (the microbial components or sites targeted by NPs are highlighted here using yellow sparks) in addition to illustrating their use for developing infection resistant PD fluid composition. In *(A)* and *(B)*, the acronyms AM, NP, NAPs and MTS represent, respectively, anti-microbial, nano-particle, nuclear associated proteins and membrane transport system.

Infectious peritonitis is not only the major cause of technique failure, but it is also the leading cause of mortality and morbidity in PD patients. Around 18% of the infection-related mortality in PD patients is the result of peritonitis [2,9,10]. Although less than 4% of peritonitis episodes result in death, it is a "contributing factor" to death in 16% of deaths on PD [9]. Statistically, about 60-70 deaths per 1000 PD patients are seen every year and the proportion is higher in developing countries where majority of the patients belong to families of low socioeconomic status [11-13]. As the number of ESRF patients continuing on PD is increasing every year (because of unavailability of kidneys required for the transplantation and various other limitations) and majority of them remain on PD for more than a year, therefore in near future, the infectious peritonitis is going to be one of the most serious healthcare associated complications [14,15]. According to an estimate, for every 0.5 increase in peritonitis rate per year, the risk of death increases by 4 % [13]. The above statistical figures, clearly highlight the public health importance of improving the existing PD technique in terms of its efficacy and adequacy to limit infections during long-term PD; so that the traumatic and life-threatening episodes of infectious peritonitis can be reduced [8,9]. Practically, this can be achieved through developing infection resistant PD fluid composition which provides long term protection against variety of PD associated infections including bacteria, mycobacteria, fungi or viruses. And, the key requisite to develop such an efficient and novel composition is that the composition should contain some antimicrobial agents with novel (non-antibiotic) mode of action to avoid the problems arising because of antibiotic resistance. Further, these antimicrobial agents should preserve their efficacy and adequacy (i.e. biocompatibility and non-cytotoxicity) during long-term intraperitoneal use.

In the above context, antimicrobial nanoparticles (AM-NPs) -especially, the metal and metal oxide NPs- would be of potential interest owing to their exclusive antimicrobial activity against variety of clinical infections [16-20] wound healing [21-23] and anti-inflammatory properties [24,25]. These NPs have been researched extensively in the past and some classes of metal (e.g. Gold, Silver, Titanium, Bismuth etc.) and metal oxide (e.g. zinc oxide, titanium oxide, etc.) nanoparticles have been found to be very effective in terms of their antimicrobial properties [26-30] and anti-biofilm properties [31,32]. Most importantly, these NPs produce their antimicrobial (AM) activity through affecting multiple biological pathways **(Fig. 1B)** found in microbial species and many concurrent mutations would have to occur in order to develop resistance against antimicrobial activity of these NPs [19]. Therefore, as required to manage recurrent and persistent infections during long-term PD, these NPs can be used frequently without posing any risk of developing antibiotic resistance. Studies have also demonstrated that naturally occurring bacteria do not develop antimicrobial resistance to metal NPs [18]. Some classes of metallic NPs have potential to eradicate multidrug-resistant infections when used in combination with antibiotics (synergic-effect) [18,19]. Like-wise, some classes can limit biofilm formation either independently [31,32] or when used in combination with antibiotics (through synergic effect) [16]. Owing to such broad spectral antimicrobial properties and activity against biofilms and formidable multidrug resistant pathogens, metallic nanoparticles are finding their potential applications as antimicrobial agents and disinfectants to improve several biomedical devices, pharmaceutical products and healthcare interventions including medicines [16,18,19,26,33].

Metallic nanoparticles of varying sizes, shapes, and properties can be synthesized using variety of chemical and physical methods [28,34-36]. However, these methods are neither cost-effective nor eco-friendly and further lead to the presence of some toxic chemicals adsorbed on the surface that could produce intolerable toxicity to humans and adverse effects in biomedical applications [34,37-40]. This is not an issue when it comes to biologically synthesized nanoparticles i.e. those synthesized from biomaterials derived from micro-organisms or plant parts following green chemistry approach [41-43]. Biological synthesis of NPs is considered as environmentally benign replacement to the toxic chemical and physical methods and studies have shown that biogenic NPs exhibit benign pharmacology response (i.e. better biocompatibility and less cytotoxicity) compared to their counterparts prepared using physical and chemical methods [44-47]. A recent study has shown that biogenic Zinc oxide NPs exhibit significantly higher biocidal activity against various pathogens when compared to chemically synthesized ZnO nanoparticles. Such preliminary studies are clearly suggesting that biogenic NPs have huge potential to address future medical concerns [34].

Biological synthesis of nanoparticles is gaining tremendous popularity owing to its cost-effectiveness, simplicity and eco-friendliness [34,37,38,40,43,48-52]. However, biological synthesis of nanoparticles using microorganisms has been found to be more tedious and time taking process as it requires more steps in maintaining cell culture, longer incubation time for the intracellular reduction of metal ions and more purification steps. On the other hand, the plant mediated green synthesis of nanoparticles is relatively simple and provides several clear advantages like (a) avoidance of maintaining time-consuming microbial cultures and purification steps, (b) rapid extracellular biosynthesis as water soluble phytochemicals reduce the metal ions in a much lesser time, (c) safe to handle and cost effective and (d) availability of broad variability of metabolites that may aid in reduction [52] [34,36-39,43,49-51]. The generalized flow chart for plant mediated nano-biosynthesis is shown in **Figure 2** schematically.

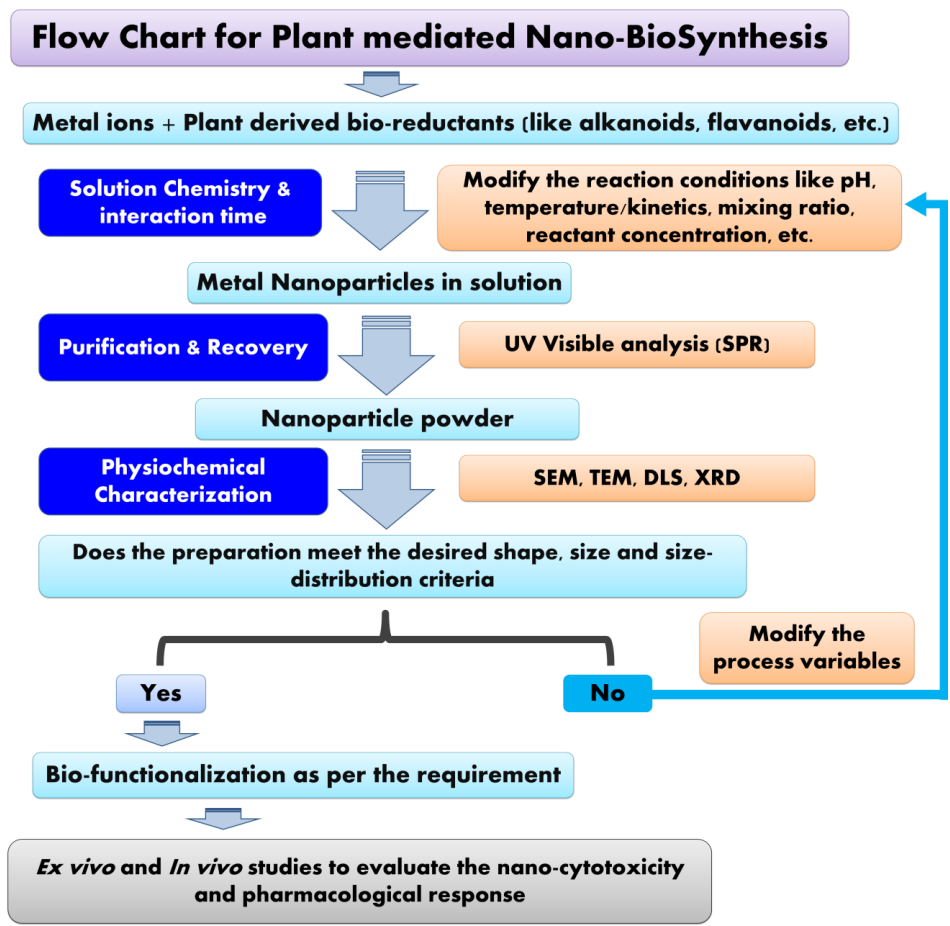

***Figure 2:*** *Flowchart denoting the* plant-mediated biological synthesis of nanoparticles.

Recently, immense applications of biogenic metal NPs have been envisaged in various biological and clinical settings [45-47,53-55]. Here, I envisage that biogenic metal NPs can also be used to attribute infection resistant properties to peritoneal dialysis fluid. However before putting these biogenic nanomaterials into human healthcare actions, the key step is to rule out their nano-toxicity and adverse effects on long-term exposure. For this, conscience efforts are required to evaluate their pharmacology through conducting dose and time of exposure dependent *ex vivo* and *in vivo* studies on human cell lines and animal models. After successful evaluation of preclinical efficacy and toxicity, the promising biogenic nanoparticles -with efficient antimicrobial properties and exhibiting favorable *in vivo* pharmacology response- can be envisaged to develop infection resistant composition of peritoneal dialysis (PD) fluid. Simply, this will be achieved through adding non-toxic doses of biogenic NPs into different types of PD fluids widely used in clinics and subsequently, evaluating (a) their efficacy against variety of healthcare associated infections, (b) *in vivo* adequacy, and (c) time stability to ensure their long-term storage and prolonged shelf life. The particular advantage of employing biogenic AM-NPs in this translational research endeavor is that these can be filter-sterilized and added directly to the PD fluid for long term storage and prolonged shelf life [56]. Further, these can easily withstand temperature variations (ranging from 4 to 50 °C which is generally encountered during transportation and storage of medical products), under which conventional antibiotics may inactivate or degrade. Moreover, the preparation of nanoparticles is cost-effective and relatively simple compared to antibiotics synthesis. These were the potential advantages which derived my interest to write this perspective to highlight the use of biogenic nanomaterials for developing infection resistant composition of PD fluid.

In conclusion, necessity to improve the PD technology for limiting frequent PD associated infections and possibility to encompass the benefits of AM-NPs synthesized biologically, have been discusses here. Particularly, the biocompatibility and cytotoxicity of biogenic AM-NPs are the key issues to be addressed before putting them into clinical applications. Further, the use of novel AM-NPs in combination with novel antibiotic agents can also be explored to manage multidrug resistant pathogens and formation of biofilms during long-term PD. The antibiotics can be added directly into the PD fluid at the time of intraperitoneal instillation or through nano-drug carriers to facilitate their slow and sustained release. I foresee that the research perspective presented here will definitely appeal some of the biomedical and clinical researchers (involved in developing, implementing, and evaluating infection prevention and control programs for healthcare settings) to put their conscience efforts in the direction to develop infection resistant PD fluid composition or nanotechnology based solutions targeting biofilms for efficient and long-term management of peritoneal dialysis. The improvisation will help the PD patients in two ways: (a) first it will provide them a better quality of life through reducing the frequency infectious peritonitis and (b) second it will reduce the financial burden on such patients through reducing medical costs and their visits to hospitals. Other major problem associated with long-term PD is the recurrent and persistent infections on the catheter exit site as it remains constantly exposed to the non-sterile environment outside the body. Therefore, efforts are also required to make the catheter exit site sterile through covering it with smart antimicrobial gels providing long-term protection without posing any risk of antibiotic resistant.

## Acknowledgement:

I would like to acknowledge Dr. Narayan Prasad (Department of Nephrology, SGPGIMS, Lucknow-226014), Dr. V Karthick (Nanoscience Division, Centre for Ocean Research, Sathyabama University, Chennai - 600119) and Dr. Sudipta Saha (Department of Pharmaceutical Sciences, Babasaheb Bhimrao Ambedkar University, Vidya Vihar, Raibareli Road, Lucknow-226025, Uttar Pradesh) for all their useful suggestions and critical comments while preparing this research perspective.